\documentclass[twocolumn]{jpsj3} 
\usepackage{bm,amssymb,amsmath}	
\usepackage{graphicx}   
\usepackage{txfonts}
\newcommand{\simg}{\stackrel{>}{_\sim}}
\newcommand{\siml}{\stackrel{<}{_\sim}}

\title{
\begin{center}
Metal-insulator Transition and Superconductivity\\ in the Two-orbital Hubbard-Holstein Model\\ for Iron-based Superconductors
\end{center}}
\author{Takemi {\sc Yamada}\thanks{E-mail address: takemi@phys.sc.niigata-u.ac.jp}, Jun {\sc Ishizuka} and Yoshiaki {\sc \=Ono}}

\inst{Department of Physics, Niigata University, Ikarashi, Nishi-ku, Niigata, 950-2181, Japan}

\abst{
We investigate a two-orbital model for iron-based superconductors 
to elucidate the effect of interplay between electron correlation and Jahn-Teller electron-phonon coupling 
by using the dynamical mean-field theory combined with the exact diagonalization method. 
When the intra- and inter-orbital Coulomb interactions, $U$ and $U'$, increase with $U=U'$, both the local spin and orbital susceptibilities, 
$\chi_{s}$ and $\chi_{o}$, increase with $\chi_{s}=\chi_{o}$ in the absence of the Hund's rule coupling $J$ and the electron-phonon coupling $g$. 
In the presence of $J$ and $g$, there are distinct two regimes: for $J \simg 2g^2/\omega_0$ with the phonon frequency $\omega_0$, 
$\chi_{s}$ is enhanced relative to $\chi_{o}$ and shows a divergence at $J=J_c$ above which the system becomes Mott insulator, 
while for $J \siml 2g^2/\omega_0$, $\chi_{o}$ is enhanced relative to $\chi_{s}$ 
and shows a divergence at $g=g_c$ above which the system becomes bipolaronic insulator. 
In the former regime, the superconductivity is mediated by antiferromagnetic fluctuations enhanced 
due to Fermi-surface nesting and is found to be largely dependent on carrier doping. 
On the other hand, in the latter regime, the superconductivity is mediated by ferro-orbital fluctuations 
and is observed for wide doping region including heavily doped case without the Fermi-surface nesting. 
}

\begin{document}
\maketitle
\section{INTRODUCTION}
The iron-based superconductors exhibit the common feature of phase diagrams, 
where parent compounds show the tetragonal-orthorhombic structural transition and the stripe-type antiferromagnetic (AFM) transition 
both of which are suppressed by carrier doping $x$ resulting in the high-$T_c$ superconductivity\cite{JACS.130.3296,AdvPhys.59.803}. 
When approaching the AFM transition, the AFM fluctuation observed by the NMR experiments\cite{PhysRevLett.104.037001} is found to be enhanced, 
while, when approaching the structural transition, 
the ferro-orbital (FO) fluctuation between $d_{xz}$ and $d_{yz}$ orbitals (or the $O_{x^2 -y^2}$ ferroquadrupole fluctuation)\cite{notation} 
responsible for the softening of the elastic constant $C_{66}$ observed 
by the ultrasonic experiments\cite{PhysRevLett.101.157003,JPSJ.81.024604,JPSJ.80.073702} is found to be enhanced. 
Correspondingly, two distinct $s$-wave pairings: the $s_{\pm}$-wave with sign change of the order parameter 
between the hole and the electron Fermi surfaces (FSs) mediated by the AFM fluctuation\cite{PhysRevLett.101.057003,PhysRevLett.101.087004} 
and the $s_{++}$-wave without the sign change mediated by the FO fluctuation\cite{JPSJ.79.123707,SSC.152.701} 
and by the antiferro-orbital (AFO) fluctuation\cite{PhysRevLett.103.177001} 
which is also responsible for the softening of $C_{66}$ through the two-orbiton process\cite{PhysRevB.84.024528}, were proposed. 

Recent experiments have revealed that the high-$T_c$ superconductivity is realized even in the case with heavily electron-doped compounds 
such as RFeAsO$_{1-x}$H$_{x}$ (R=Sm, Ce, La)\cite{PhysRevB.84.024521,PhysRevB.85.014514,NatureComm.3.943} up to $x\sim 0.5$ 
and A$_x$Fe$_2$Se$_2$ (A=K, Cs, Rb)\cite{PhysRevB.82.180520,NatureMat.10.273} where the large electron FSs are observed without the hole FSs. 
In these cases, the mechanisms based on the AFM\cite{PhysRevLett.101.057003,PhysRevLett.101.087004} 
and the AFO\cite{PhysRevLett.103.177001} fluctuations, 
which are enhanced due to the nesting between the electron and hole FSs, seem to be insufficient for explaining the superconductivity. 
Therefore, the other types of the nesting between the electron FSs for A$_x$Fe$_2$Se$_2$\cite{PhysRevB.83.100515,PhysRevB.83.140512} 
and those due to the effects of the doping dependence of the band structure for RFeAsO$_{1-x}$H$_{x}$\cite{JPSJ.82.083702,PhysRevB.88.041106} have been discussed. 
As for the mechanism based on the FO fluctuation\cite{JPSJ.79.123707,SSC.152.701} 
which is enhanced due to the coupling between the $d_{xz}-d_{yz}$ orbital fluctuation and the orthorhombic mode (Jahn-Teller type) phonon, 
the superconductivity does not need the FS nesting effect 
but is restricted near the tetragonal-orthorhombic structural transition with small $x$ within the random phase approximation (RPA). 

Generally, in the magnetic fluctuation mechanism, the the pairing interaction $V(\bm{q})$ with wave vector $\bm{q}$ is repulsive 
and then the strong $\bm{q}$ dependence of $V(\bm{q})$ realized near the magnetic ordered phase is crucial for the superconductivity. 
On the other hand, in the orbital fluctuation mechanism, 
$V(\bm{q})$ is attractive and then the strong $\bm{q}$ dependence of $V(\bm{q})$ realized near the orbital ordered phase is not necessary for the superconductivity. 
When the local component of the orbital fluctuation is relatively larger than that of the magnetic fluctuation, 
the local component of the pairing interaction $V_{\rm loc}$, which is nothing but the $\bm{q}$-averaged value of $V(\bm{q})$, 
becomes attractive and is expected to induce the $s_{++}$-wave pairing, even far away from the ordered phases. 
It has been found that the local magnetic (charge) susceptibility is largely enhanced 
due to the effect of strong electron correlation (strong electron-phonon coupling) 
and shows a divergence towards the Mott (bipolaronic) metal-insulator transition\cite{RevModPhys.68.13,PhysRevB.70.155103,EurophysLett.66.559,PhysRevB.74.205108}. 
Therefore, it is important in describing the local fluctuations to take into account of both the strong correlation 
and coupling effects beyond perturbative approaches such as the RPA.

In this paper, we investigate the effect of interplay between electron correlation and Jahn-Teller (JT) electron-phonon coupling 
including the strong correlation and/or strong coupling regimes by using the dynamical mean-field theory (DMFT)\cite{RevModPhys.68.13,JPSJ.82.123712} 
which becomes exact in infinite dimensions ($d=\infty$) where the self-energy becomes local 
and enables us to sufficiently include the local correlation effects due to both Coulomb and electron-phonon interactions 
and to describe the Mott and bipolaronic metal-insulator transitions\cite{PhysRevB.70.155103,EurophysLett.66.559,PhysRevB.74.205108,PhysC.426.330,JPSJ.79.054707}. 
Here, we employ the two orbital Hubbard model\cite{PhysRevB.77.220503} reproducing the electron and hole FSs as a minimal model for iron-based superconductors. 
The model has been extensively studied by many authors focusing on 
the pairing states\cite{FrontPhys.6.379,JSNM.22.539,NJP.11.025009,PhysC.469.932,PhysC.471.453,PhysRevB.78.195114,PhysRevB.79.014508,PhysRevB.79.134502,PhysRevB.81.014511,PhysRevB.81.104504,PhysRevB.81.144509,PhysRevB.85.024532,PhysRevB.81.144521,PhysRevLett.101.237004,PhysRevLett.102.047006,PhysRevLett.106.217002,JPCS.400.022134}, 
magnetic states\cite{JPCM.23.246001,JPCM.23.312201,JPSJ.78.083704,PhysRevB.79.235207,PhysRevB.81.085106,PhysRevB.81.172504,PhysRevB.82.045125,PhysRevB.82.104520,PhysRevB.83.174513,PhysRevB.83.224503,PhysRevB.84.094519,PhysRevB.84.174512,PhysRevB.85.035123,PhysRevB.85.214425}, 
strong correlation effects\cite{PhysRevB.79.064517,PhysRevB.84.235115,PhysRevLett.106.186401,EPL.95.17003} 
and lattice and orbital properties\cite{NJP.11.013051,PhysRevB.83.092505,PhysRevB.84.064435}. 
However, the effect of the JT phonon, which has been found to play important role in the orbital fluctuations 
responsible for the softening of $C_{66}$\cite{JPSJ.79.123707,SSC.152.701}, was not discussed there. 
The purpose of this paper is to elucidate the effect of the JT electron-phonon coupling in the two-orbital Hubbard model 
including the strong correlation and/or strong coupling regimes which were not considered in the previous works based on the RPA\cite{JPSJ.79.123707,SSC.152.701}.  

\section{MODEL AND FORMULATION}
\subsection{Model Hamiltonian}
Our model Hamiltonian is given by 
\begin{align}
H=H_{\rm 0}+H_{\rm int}+H_{\rm ph}+H_{\rm el-ph} 
\label{eq:H} 
\end{align}
with the kinetic part of the Hamiltonian: 
\begin{align}
 &\!\!\!H_{\rm 0}=\!\!\sum_{\bm{k}\sigma}
\left(d_{\bm{k}1\sigma}^{\dagger}~\!d_{\bm{k}2\sigma}^{\dagger}\right)
\!\hat{H_{\bm{k}}}\!
\left(\begin{array}{l}
\!\!d_{\bm{k}1\sigma}\!\!\!\\
\!\!d_{\bm{k}2\sigma}\!\!\!\\
\end{array}\right)\!, \ \ 
\!\!\hat{H_{\bm{k}}}\!=\!\left(\begin{array}{ll}
\!\!\!\varepsilon_{\bm{k}1} \!   &\!\varepsilon_{\bm{k}12}\!\!\! \\
\!\!\!\varepsilon_{\bm{k}12}\!\! &\!\varepsilon_{\bm{k}2}\!\! \\
\end{array}\right)\!
\label{eq:H0}
\end{align}
where $d_{\bm{k}l\sigma}$ is the annihilation operator for a Fe $3d$ electron with the wave vector $\bm{k}$ 
and the spin $\sigma$ in the orbital $l$=$1,2$ (=$d_{xz},d_{yz}$), 
and the energies $\varepsilon_{\bm{k}l}$ and $\varepsilon_{\bm{k}12}$ are determined 
so as to reproduce the electron and hole FSs in the iron pnictides\cite{PhysRevB.77.220503}. 
The Coulomb interaction part $H_{\rm int}$ between electrons at site $i$ 
includes the intra- and inter-orbital direct terms $U$ and $U'$, the Hund's rule coupling $J$ 
and the pair transfer $J'$ which is written by,
\begin{align}
H_{\rm int}
&=U\sum_{il}n_{il\uparrow}n_{il\downarrow}+\frac{U'}{2}\sum_{i}\sum_{l\neq l'}\sum_{\sigma\sigma'}n_{il\sigma}n_{il'\sigma'} \nonumber \\
&+\frac{J}{2}\sum_{i}\sum_{l\neq l'}\sum_{\sigma\sigma'}d_{il\sigma}^{\dagger}d_{il'\sigma'}^{\dagger}d_{il\sigma'}d_{il'\sigma} \nonumber \\
&+\frac{J'}{2}\sum_{i}\sum_{l\neq l'}\sum_{\sigma\neq\sigma'}d_{il\sigma}^{\dagger}d_{il\sigma'}^{\dagger}d_{il'\sigma'}d_{il'\sigma}.
\label{eq:Hint}
\end{align} 
For simplicity, we assume the relations $U$=$U'$+$2J$ and $J$=$J'$ 
which are satisfied in the isolated atom but not generally in the crystal\cite{JPSJ.77.123701,PhysRevB.81.054518}. 
The phonon and the electron-phonon interaction parts are given by 
\begin{align}
&H_{\rm ph}+H_{\rm el-ph}=\sum_{i}\omega_{0}b_{i}^{\dagger}b_{i}+g\sum_{i}\left(b_{i}+b_{i}^{\dagger}\right)\tau_{zi},
\label{eq:Hph}
\end{align}
where $b_{i}$ is the annihilation operator for a JT phonon at site $i$ with the frequency $\omega_{0}$, 
which is coupled to the longitudinal orbital fluctuation, 
$\tau_{zi}$=$\sum_{\sigma}(n_{i1\sigma}-n_{i2\sigma})$ with $n_{il\sigma}$=$d_{il\sigma}^{\dagger}d_{il\sigma}$, 
through the electron-phonon coupling $g$. 

\subsection{Formulation : Dynamical mean-field theory}
To solve the model Eq. (\ref{eq:H}), we use the DMFT\cite{RevModPhys.68.13,JPSJ.82.123712} in 
which the lattice model is mapped onto an impurity Anderson model embedded in an effective medium which is
determined so as to satisfy the self-consistency condition
\begin{eqnarray}
{\hat G}(i\varepsilon_n)=\frac{1}{N}\sum_{\bm{k}}{\hat G}({\bm k},i\varepsilon_n)
\label{eq:scf}
\end{eqnarray}
with the Matsubara frequency $\varepsilon_n$=(2$n$+1)$\pi T$, 
where ${\hat G}(i\varepsilon_n)$ and ${\hat G}({\bm k},i\varepsilon_n)$ are the 2$\times$2 matrix representations
of the local (impurity) Green's function and the lattice Green's function, respectively, 
which are explicitly given by
\begin{align}
{\hat G}(i\varepsilon_n)&=\left[[{\cal \hat{G}}(i\varepsilon_n)]^{-1} 
-{\hat \Sigma}(i\varepsilon_n)\right]^{-1},
\label{eq:G_loc}\\
{\hat G}({\bm k},i\varepsilon_n)&=\left[(i\varepsilon_n+\mu)-{\hat H}_{{\bm k}}
-{\hat \Sigma}(i\varepsilon_n)\right]^{-1}, 
\label{eq:G}
\end{align} 
where ${\hat \Sigma}(i\varepsilon_n)$ is the $2\times2$ matrix representation 
of the impurity (local) self-energy and 
${\cal\hat{G}}(i\varepsilon_n)$ is that of the bare impurity Green's function 
describing the effective medium which is determined self-consistently.

Within the DMFT, the spin (charge-orbital) susceptibility\cite{JPSJ.82.123712} is given in the $4\times4$ matrix representation as 
\begin{align}
\hat{\chi}_{s(c)}(q)=
\left[1 -(+)\hat{\chi}_0(q)\hat{\Gamma}_{s(c)}(i\omega_n)\right]^{-1} \hat{\chi}_0(q)
\label{eq:chi}
\end{align}
with 
\begin{eqnarray}
\hat{\chi}_0(q)=-\frac{T}{N}\sum_{k}\hat{G}(k+q)\hat{G}(k),
\end{eqnarray}
where $k$=$(\bm{k},i\varepsilon_n)$, $q$=$(\bm{q},i\omega_m)$ and $\omega_m$=$2m\pi T$.  
In Eq. (\ref{eq:chi}), $\hat{\Gamma}_{s(c)}(i\omega_m)$ is the local irreducible spin (charge-orbital) vertex 
in which only the external frequency ($\omega_m$) dependence is considered as a simplified approximation\cite{zdependence} and is explicitly given by
\begin{eqnarray}
\hat{\Gamma}_{s(c)}(i\omega_m)=-(+)\left[\hat{\chi}_{s(c)}^{-1}(i\omega_m)-\hat{\chi}_0^{-1}(i\omega_m)\right]
\label{eq:gamma}
\end{eqnarray} 
with
\begin{eqnarray}
\hat{\chi}_0(i\omega_m)=-T\sum_{\varepsilon_n}
\hat{G}(i\varepsilon_n+i\omega_m)\hat{G}(i\varepsilon_n), 
\end{eqnarray} 
where $\hat{\chi}_{s(c)}(i\omega_m)$ is 
the local spin (charge-orbital) susceptibility corresponding to the $\bm{q}$-averaged value of $\hat{\chi}_{s(c)}(q)$ in eq. (\ref{eq:chi}) 
and is explicitly defined by 
\begin{align}
&\left[\hat{\chi}_{s(c)}(i\omega_m)\right]_{l_1l_2l_3l_4}\nonumber\\
&=\sum_{\sigma\sigma'}A_{s(c)}^{\sigma\sigma'}
\int_{0}^{\beta}\!\!d\tau e^{i\omega_m\tau}
\langle d_{il_1\sigma}^{\dagger}(\tau)d_{il_2\sigma'}(\tau)d_{il_4\sigma'}^{\dagger}(0)d_{il_3\sigma}(0)\rangle 
\label{eq:chiloc}
\end{align}
with $A_{s(c)}^{\uparrow\uparrow}$=$A_{s(c)}^{\downarrow\downarrow}$=1 
and $A_{s(c)}^{\uparrow\downarrow}$=$A_{s(c)}^{\downarrow\uparrow}$=$-1(1)$. 
When the largest eigenvalue $\alpha_s$ ($\alpha_c$) of  $(-)\hat{\chi}_0(q)\hat{\Gamma}_{s(c)}(i\omega_m)$ in Eq. (\ref{eq:chi}) 
for a wave vector $\bm{q}$ with $i\omega_m$=$0$ reaches unity, 
the instability towards the magnetic (charge-orbital) order with the corresponding $\bm{q}$ takes place\cite{gamma}.

To examine the superconductivity mediated by the magnetic and charge-orbital fluctuations which are enhanced towards the corresponding orders mentioned above, 
we write the effective pairing interaction for the spin-singlet state using the spin (charge-orbital) susceptibility 
and vertex given in Eqs. (\ref{eq:chi}) and (\ref{eq:gamma}) obtained 
within the DMFT in the $4\times4$ matrix representation as\cite{JPSJ.79.123707,PhysRevB.82.064518,PhysRevB.69.104504} 
\begin{align}
\hat{V}(q)
&=\frac{3}{2}\hat{\Gamma}_{s}(i\omega_m)\hat{\chi}_{s}(q)\hat{\Gamma}_{s}(i\omega_m)
+\frac{1}{2}\hat{\Gamma}_{s}^{(0)}\nonumber\\
&-\frac{1}{2}\hat{\Gamma}_{c}(i\omega_m)\hat{\chi}_{c}(q)\hat{\Gamma}_{c}(i\omega_m)
+\frac{1}{2}\hat{\Gamma}_{c}^{(0)}(i\omega_m),
\label{eq:pair}
\end{align}
where the bare spin (charge-orbital) vertex is given by 
\begin{equation}
\Gamma_{s(c)}^{(0)}= \left\{
\begin{array}{@{\,} l @{\,} c}
U~(U-2g^2 D(i\omega_m)) & (l_1=l_2=l_3=l_4)\\
U'~(-U'+2J) & (l_1=l_3\ne l_2=l_4)\\
J~(2U'-J+2g^2 D(i\omega_m)) & (l_1=l_2\ne l_3=l_4)\\
J'~(J')& (l_1=l_4\ne l_2=l_3)\\
0 & (\mathrm{otherwise})
\end{array} \right.  
\label{eq_U}
\end{equation} 
with the bare phonon Green's function\cite{JPSJ.79.123707,PhysRevB.82.064518} 
\begin{equation}
D(i\omega_m)=\frac{2\omega_0}{\omega_m^2+\omega_0^2}.
\end{equation}
Substituting the effective pairing interaction Eq. (\ref{eq:pair}) into 
the linearized Eliashberg equation: 
\begin{align}
\lambda \Delta_{ll'}(k)&=-\frac{T}{N}\sum_{k'}
\sum_{l_1l_2l_3l_4}V_{ll_1,l_2l'}(k-k')\ \ \ \ \ \ \nonumber\\
&\times G_{l_3l_1}(-k')
\Delta_{l_3l_4}(k') G_{l_4l_2}(k'), 
\label{gapeq}
\end{align}
we obtain the gap function $\Delta_{ll'}(k)$ with the eigenvalue $\lambda$ 
which becomes unity at the superconducting transition temperature $T$=$T_c$.  
In Eq. (\ref{gapeq}), 
the gap function $\Delta_{ll'}(k)$ includes the $1/d$ corrections yielding 
the ${\bm k}$ dependence of the gap function responsible for the anisotropic superconductivity 
which is not obtained within the zeroth order of $1/d$\cite{RevModPhys.68.13}. 
If we replace $\hat{\Gamma}_{s(c)}$ with $\hat{\Gamma}_{s}^{(0)}$ and neglect $\hat{\Sigma}$, 
Eq. (\ref{eq:pair}) yields the RPA result of $\hat{V}(q)$\cite{PhysRevB.82.064518,JPSJ.77.123701,PhysRevB.81.054518,PhysRevB.69.104504}. 
Therefore, Eq. (\ref{gapeq}) with Eqs. (\ref{eq:G}) and (\ref{eq:pair}) is a straightforward extension 
of the RPA result to include the vertex and the self-energy corrections within the DMFT without any double counting.\cite{JPSJ.82.123712} 

In the actual calculations with the DMFT, 
we solve the effective impurity Anderson model, 
where the Coulomb and the JT electron-phonon interactions at the impurity site are given by the same forms as Eqs. (\ref{eq:Hint}) and (\ref{eq:Hph}) 
with a site $i$ and the kinetic energies are determined so as to satisfy Eq. (\ref{eq:scf}) as possible, 
by using the exact diagonalization (ED) method for a finite-size cluster to obtain the local quantities such as $\hat{\Sigma}$ and $\hat{\chi}_{s(c)}$. 
We set the site number $N_s$=$4$-$6$ and the cutoff of the phonon number $N_{b}$=$20$\cite{PhysC.426.330,JPSJ.79.054707}. 
The tight-binding parameters of $H_{\rm k}$ in Eq. (\ref{eq:H0}) are set to be the same in Ref.\cite{PhysRevB.77.220503} 
where the total band width is $W=12$ with the nearest neighbor transfer $t=1$ 
which corresponds to $W\sim 4$ eV ($t\sim0.33$ eV) in the typical 
$d$-band width of the iron-based superconductor, 
and we set the phonon frequency $\omega_{0}$=$0.01W$. 
All calculations are performed at $T$=$0$, and we replace the Matsubara frequencies $\varepsilon_n$ and $\omega_m$ 
by a fine grid of imaginary frequencies with a fictitious temperature which determines the energy resolution. 

Using the ED method, we also calculate the several physical quantities as follows: 
the renormalization factor corresponding to the inverse effective mass 
$Z$=$(1-\frac{\rm{d} \Sigma(i\varepsilon)}{\rm{d} (i\varepsilon)}|_{_{i\varepsilon=0}})^{-1}$=$(m^*/m)^{-1}$, 
the local charge, spin and orbital fluctuations 
$\langle\delta n^{2}\rangle$=$\langle(n-\langle n\rangle)^2\rangle$ with $n$=$\sum_{l\sigma}n_{i1\sigma}$, 
$\langle\bm{S}^{2}\rangle$ with $\bm{S}$=$\frac{1}{2}\sum_{l}\sum_{\alpha\beta}d_{il\alpha}^{\dagger}\bm{\sigma}_{\alpha\beta}d_{il\beta}$ and 
$\langle\tau_{z}^{2}\rangle$, 
and the local spin and orbital susceptibilities 
$\chi_{s}$=$4\langle \langle S_{z}|S_{z} \rangle \rangle|_{i\omega=0}$ and 
$\chi_{o}$=$\langle \langle \tau_{z}|\tau_{z} \rangle \rangle|_{i\omega=0}$, 
and the intra-orbital part of the local paring interaction 
$V_{\rm loc}\equiv[\hat{V}(0)]_{llll}$ with $\hat{V}(i\omega)$=
$\frac{1}{N}\sum_{\bm{q}}\hat{V}(\bm{q},i\omega)$=
$\frac{3}{2}\hat{\Gamma}_{s}(i\omega)\hat{\chi}_{s}(i\omega)\hat{\Gamma}_{s}(i\omega)
-\frac{1}{2}\hat{\Gamma}_{c}(i\omega)\hat{\chi}_{c}(i\omega)\hat{\Gamma}_{c}(i\omega)
+\frac{1}{2}(\hat{\Gamma}_{s}^{(0)}+\hat{\Gamma}_{c}^{(0)}(i\omega))$
which is the most dominant contribution of the pairing interaction due to the local fluctuations.

\section{RESULTS}
\subsection{Effects of Coulomb interactions with $U=U'$}

\begin{figure}[t]
\begin{center}
\includegraphics[width=7.0cm]{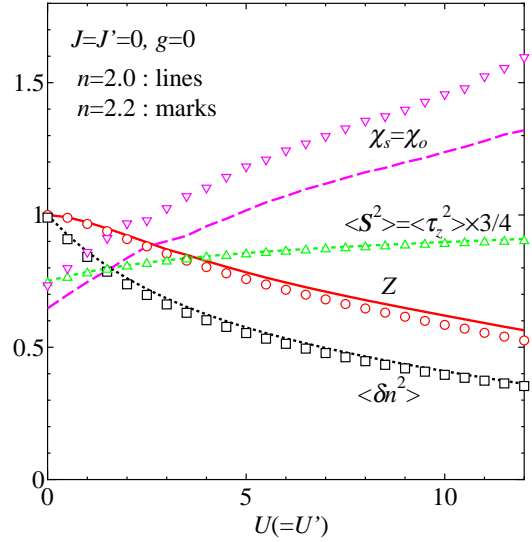}
\caption{(Color online) 
$U$ ($=U'$) dependence of the renormalization factor $Z$, the local spin, orbital and charge fluctuations $\langle\bm{S}^{2}\rangle$, $\langle\delta n^{2}\rangle$, 
and the local spin and orbital susceptibilities $\chi_{s}$ and $\chi_{o}$ with $J$=$J'$=$g$=0 for $n$=$2$ (lines) and $n$=$2.2$ (marks).}
\label{fig:Fig1}
\end{center}
\end{figure}

First, we examine the effect of the intra- and inter-orbital Coulomb interactions, $U$ and $U'$ 
in the absence of the Hund's rule coupling $J$ and the JT electron-phonon coupling $g$. 
Fig. \ref{fig:Fig1} shows the several physical quantities mentioned in $\S$ 2 as functions of $U$ ($=U'$) with $J$=$J'$=$g$=$0$ 
at half-filling $n$=$2$ and away from half-filling $n$=$2.2$.  
When the electron correlation increases with $U$=$U'$, $Z$ and $\langle \delta n^{2}\rangle$ decrease 
while $\langle\bm{S}^{2}\rangle$ and $\langle\tau_{z}^{2}\rangle$ increase with $\langle\bm{S}^{2}\rangle$=$\frac{3}{4}\langle\tau_{z}^{2}\rangle$ 
as the double-occupancy probabilities take the same value: 
$\langle n_{l\uparrow}n_{l\downarrow}\rangle$=$\langle n_{l\uparrow}n_{l'\downarrow}\rangle$=$\langle n_{l\uparrow}n_{l'\uparrow}\rangle$ with $l\neq l'$ 
because of the spin-orbital symmetry\cite{JPSJ.79.054707}. 
Correspondingly, $\chi_{s}$ and $\chi_{o}$ increase with $\chi_{s}$=$\chi_{o}$\cite{chi} while the charge susceptibility decreases (not shown) with increasing $U$. 
For $n$=$2$, we also observe the Mott metal-insulator transition at a critical interaction $U_c(=U'_c)\sim 2.5W$, 
where $Z$=$0$ for $U>U_c$, while when $U \to U_c$ for $U<U_c$, $Z\to 0$ and $\chi_{s}$=$\chi_{o}\to \infty$ (not shown), 
as previously observed in the multi-orbital Hubbard model\cite{PhysRevB67.035119}. 
For $n$=$2.2$, the $U$ dependence of the physical quantities is almost similar to that for $n$=$2$ as shown in Fig. \ref{fig:Fig1}, 
except for the Mott transition which is observed exclusively for integer fillings\cite{PhysRevB67.035119}. 
We note that, although the $\bm{q}$ dependence of $\hat{\chi}_{s}(\bm{q},i\omega)$ largely depends on doping responsible for the FS nesting as will be shown later, 
the $\bm{q}$-averaged value, i. e., the local susceptibility $\hat{\chi}_{s}(i\omega)$ is weakly dependent on doping. 

\begin{figure}[t]
\begin{center}
\includegraphics[width=7.0cm]{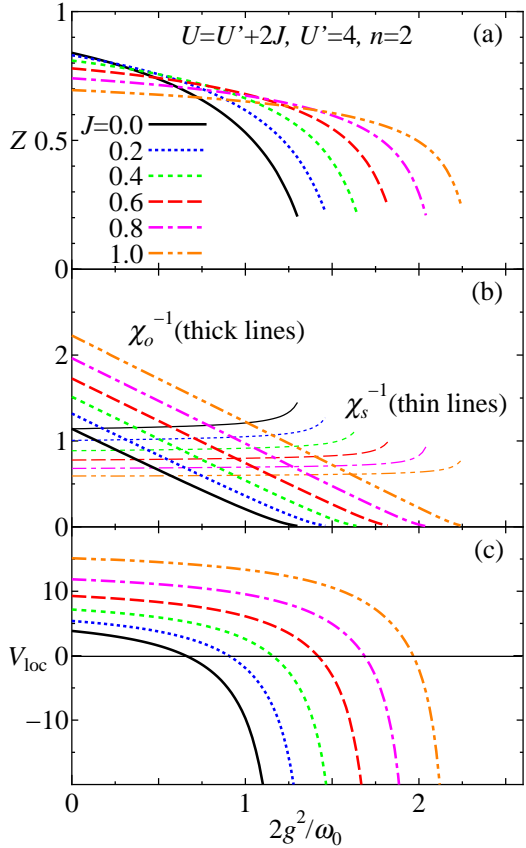}
\caption{(Color online) 
$2g^{2}/\omega_{0}$ dependence of the renormalization factor $Z$ (a), the inverse of the local spin (orbital) susceptibility $\chi_{s(o)}^{-1}$ (b) 
and the intra-orbital part of the local paring interaction 
$V_{\rm loc}$ (c) for several values of $J$(=$J'$) with $U'$=$4$ and $n$=$2$. 
}\label{fig:Fig2}
\end{center}
\end{figure}

\subsection{Effects of Hund's rule coupling $J$ and JT electron-phonon coupling $g$}
\begin{figure}[t]
\begin{center}
\includegraphics[width=7.0cm]{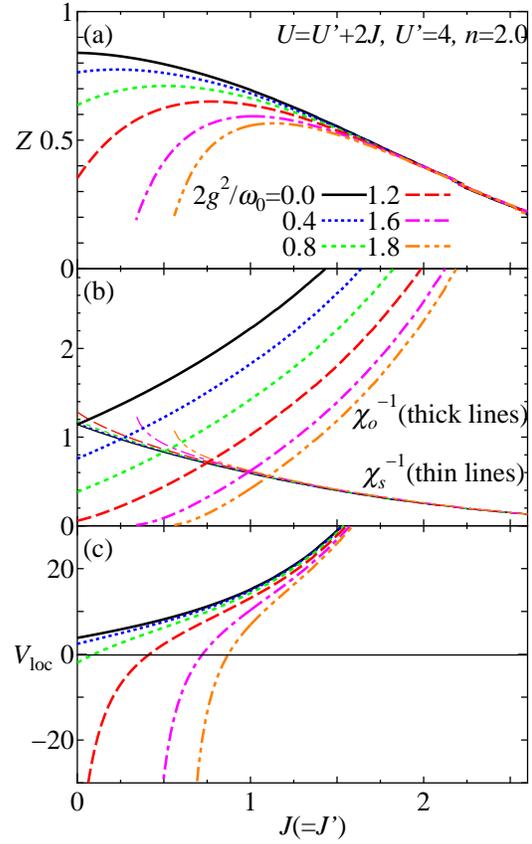}
\caption{(Color online) 
$J$(=$J'$) dependence of the renormalization factor $Z$ (a), the inverse of the local spin (orbital) susceptibility $\chi_{s(o)}^{-1}$ (b) 
and the intra-orbital part of the local paring interaction $V_{\rm loc}$ (c) for several values of $2g^{2}/\omega_{0}$ with $U'$=$4$ and $n$=$2$. 
}\label{fig:Fig3}
\end{center}
\end{figure}

Next, we consider the effects of the Hund's rule coupling $J$ and the JT electron-phonon coupling $g$. 
In Figs. \ref{fig:Fig2} (a)-(c), $Z$, $\chi_{s(o)}^{-1}$ and $V_{\rm loc}$ are plotted as functions of $2g^{2}/\omega_{0}$ 
for several values of $J$(=$J'$) for $U'$=$4$ with $U$=$U'$+$2J$ and $n$=$2$. 
When $2g^{2}/\omega_{0}$ increases, $Z$ decreases with increasing $\chi_{o}$ due to the strong orbital-lattice coupling effect, while $\chi_{s}$ slightly decreases. 
Correspondingly, $V_{\rm loc}$ decreases with increasing $2g^{2}/\omega_{0}$ and finally becomes negative 
where the attractive term due to $\chi_{o}$ dominates over the repulsive 
term due to $\chi_{s}$ (see Eq. (\ref{eq:pair})). 
Then, the intra-orbital $s$-wave pairing is expected to be realized in the intermediate coupling regime 
where $V_{\rm loc}<0$ with the moderate effective mass $m^*/m$=$Z^{-1}$=$2$. 
In the strong coupling regime, we also observe the bipolaronic transition at a critical coupling $g_c$ 
where $Z\to 0$ with $g\to g_c$ together with $\chi_{o}\to \infty$ and $V_{\rm loc}\to -\infty$, 
although it is difficult to obtain a fully convergent solution with $Z\sim 0$. 
The effect of $J$ enhances $\chi_{s}$ while suppresses $\chi_{o}$ (see also Fig. \ref{fig:Fig3}). 

Figs. \ref{fig:Fig3} (a)-(c) show the $J$ dependence of $Z$, $\chi_{s(o)}^{-1}$ and $V_{\rm loc}$ 
for several values of $2g^{2}/\omega_{0}$ for $U'$=$4$ with $U$=$U'$+$2J$ and $n$=$2$. 
When $J$ increases, $Z$ monotonically decreases for $2g^{2}/\omega_{0}$=$0$ 
while it shows a maximum at $J\sim 2g^{2}/\omega_{0}$ for $2g^{2}/\omega_{0}\ne 0$. 
$\chi_{s}(\chi_{o})$ increases (decreases) with increasing $J$ resulting in a crossover 
between the following two regimes: $J\siml 2g^{2}/\omega_{0}$ with $\chi_{s}<\chi_{o}$ and $J\simg 2g^{2}/\omega_{0}$ with $\chi_{s}>\chi_{o}$. 
Then, the large effective mass $m^*/m$=$Z^{-1}\gg 1$ is observed in the two distinct regimes with $J\ll 2g^{2}/\omega_{0}$ ($J\gg 2g^{2}/\omega_{0}$) 
where $\chi_{o}(\chi_{s})$ dominates over $\chi_{s}(\chi_{o})$ due to the strong coupling (correlation) effect, 
while the moderate effective mass $m^*/m$=$Z^{-1}\sim 2$ is observed in the intermediate regime with $J\sim 2g^{2}/\omega_{0}$ 
where $\chi_{s}$ and $\chi_{o}$, both of which are largely enhanced by $U$(=$U'$) as shown in Fig. \ref{fig:Fig1}, 
compete to each other resulting in a maximum of $Z$ as a fully non-perturbative effect. 
This intermediate regime with $\chi_{o}$ being a little larger than $\chi_{s}$ is responsible for the $s$-wave pairing 
due to $V_{\rm loc}<0$ with the moderate renormalization of the band width $Z\sim 1/2$ 
and seems to be potentially relevant for the description of  the iron-pnictide superconductors 
where both the spin and orbital fluctuations are largely enhanced 
while the band renormalization is moderate\cite{AdvPhys.59.803}.

\subsection{Phase diagrams on $J$-$2g^{2}/\omega_0$ plane}
\begin{figure}[t]
\begin{center}
\includegraphics[width=7.7cm]{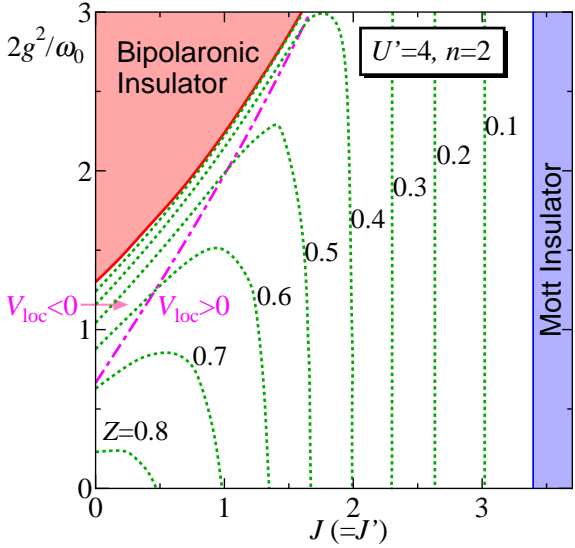}
\caption{(Color online) 
The contour lines of the renormalization factor $Z$ as functions of $J$(=$J'$) and $2g^{2}/\omega_{0}$ for $U'$=$4$ and $n$=$2$. 
The phase boundary towards the Mott and the bipolaronic insulators are the contour lines with $Z$=$0$. 
The dash-dotted line shows the boundary between the regions of $V_{\rm loc}>0$ and $V_{\rm loc}<0$. 
}\label{fig:Fig4}
\end{center}
\end{figure}

From systematic calculations for various values of $J$ and $g$, 
we obtain the contour lines of the renormalization factor $Z$ as functions of $J$(=$J'$) and $2g^{2}/\omega_{0}$ for $U'$=$4$ 
with $U$=$U'$+$2J$ and $n$=$2$ as shown in Fig. \ref{fig:Fig4}, 
where the contour lines with $Z$=$0$ are the phase boundary towards the Mott and the bipolaronic insulators. 
As mentioned before there are distinct two regimes: $J\siml 2g^{2}/\omega_{0}$ with $\chi_{s}<\chi_{o}$ and $J\simg 2g^{2}/\omega_{0}$ with $\chi_{s}>\chi_{o}$. 
In the former regime, $Z$ decreases with increasing $g$ together with increasing $\chi_{o}$ 
and then $Z\to 0$ with $\chi_{o}\to \infty$ at $g$=$g_c$ above which the system becomes bipolaronic insulator (see also Fig. \ref{fig:Fig2}), 
while $Z$ increases with increasing $J$. 
In this regime, the attractive local pairing interaction due to $\chi_o$ dominates over the repulsive one due to $\chi_s$ resulting in $V_{\rm loc}<0$. 
On the other hand, in the latter regime, $Z$ decreases with increasing $J$ together with increasing $\chi_{s}$ 
and then $Z\to 0$ with $\chi_{s}\to \infty$ at $J$=$J_c$ above which the system becomes Mott insulator (see also Fig. \ref{fig:Fig3}), 
while $Z$ is almost independent of $g$. 
In the crossover regime with $J \sim 2g^2/\omega_0$ where $\chi_{s}\sim \chi_{o}$, 
the effects of both spin and orbital fluctuations on the band renormalization compete with each other 
and then we observe a ridge of $Z$ as shown in Fig. \ref{fig:Fig4}. 

\begin{figure}[t]
\begin{center}
\includegraphics[width=7.5cm]{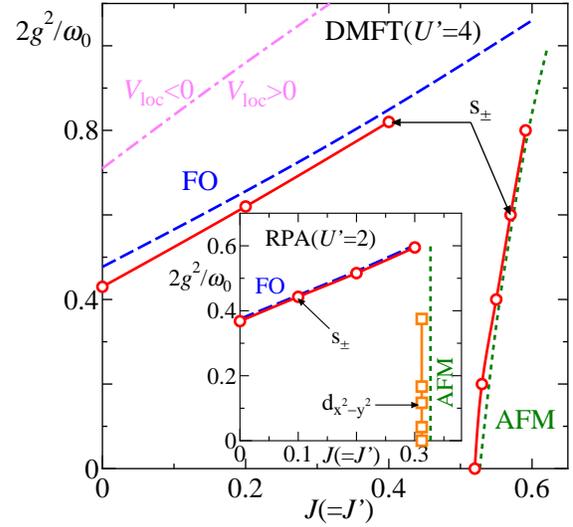}
\caption{(Color online) 
The phase diagram on $J$-$2g^{2}/\omega_{0}$ plane for $n$=$2$. 
The lines represent the instabilities for the ferro-orbital order (dashed lines), the stripe-type AFM order (dotted lines), 
the $s_{++}$-wave superconductivity (solid line with open circles) and the $d_{x^2-y^2}$-superconductivity (solid lines with open squares), respectively. 
The dash-dotted line shows the boundary between  the regions of $V_{\rm loc}>0$ and $V_{\rm loc}<0$. 
The DMFT results for $U'$=$4$ are shown in the main figure and the RPA results for $U'$=$2$\cite{uvalue} are shown in the inset. 
}\label{fig:Fig5}
\end{center}
\end{figure}

Finally, we discuss the magnetic and orbital orders and the superconductivity mediated by those fluctuations. 
Fig. \ref{fig:Fig5} shows the phase diagram on $J$-$2g^{2}/\omega_{0}$ plane for $U'$=$4$ and $n$=$2$, 
where the magnetic (charge-orbital) instability takes place 
when the largest eigenvalue $\alpha_s$ ($\alpha_c$) of $(-)\hat{\chi}_0(q)\hat{\Gamma}_{s(c)}(i\omega_m)$ in Eq. (\ref{eq:chi}) 
with $i\omega_m$=$0$ reaches unity, and the superconducting instability occurs when the largest eigenvalue $\lambda$ 
in the linearized Eliashberg equation Eq. (\ref{gapeq}) becomes unity. 
The stripe-type AFM order with $\bm{q}$=$(\pi,0)$ appears in the large $J$ region, 
while, the FO order with $\bm{q}$=$(0,0)$ appears in the large $2g^{2}/\omega_{0}$ region. 
It is noted that on the phase boundary towards the charge-orbital instability, 
the longitudinal orbital susceptibility $[\hat{\chi}_c(\bm{q})]_{11,11}-[\hat{\chi}_c(\bm{q})]_{11,22}$ diverges, 
while, the charge susceptibility $[\hat{\chi}_c(\bm{q})]_{11,11}+[\hat{\chi}_c(\bm{q})]_{11,22}$ does not. 
In the inset in Fig. \ref{fig:Fig5}, we also show the RPA phase diagram for $U'$=$2$\cite{uvalue} and $n$=$2$ for comparison. 
The AFM order from the DMFT is largely suppressed as compared to the RPA result due to the correlation effect, while the FO order is not so. 
Then, the FO order is stabilized relative to the AFM order due to the correlation effect beyond the RPA. 

\subsection{Superconductivity}
\begin{figure}[t]
\begin{center}
\includegraphics[width=8.0cm]{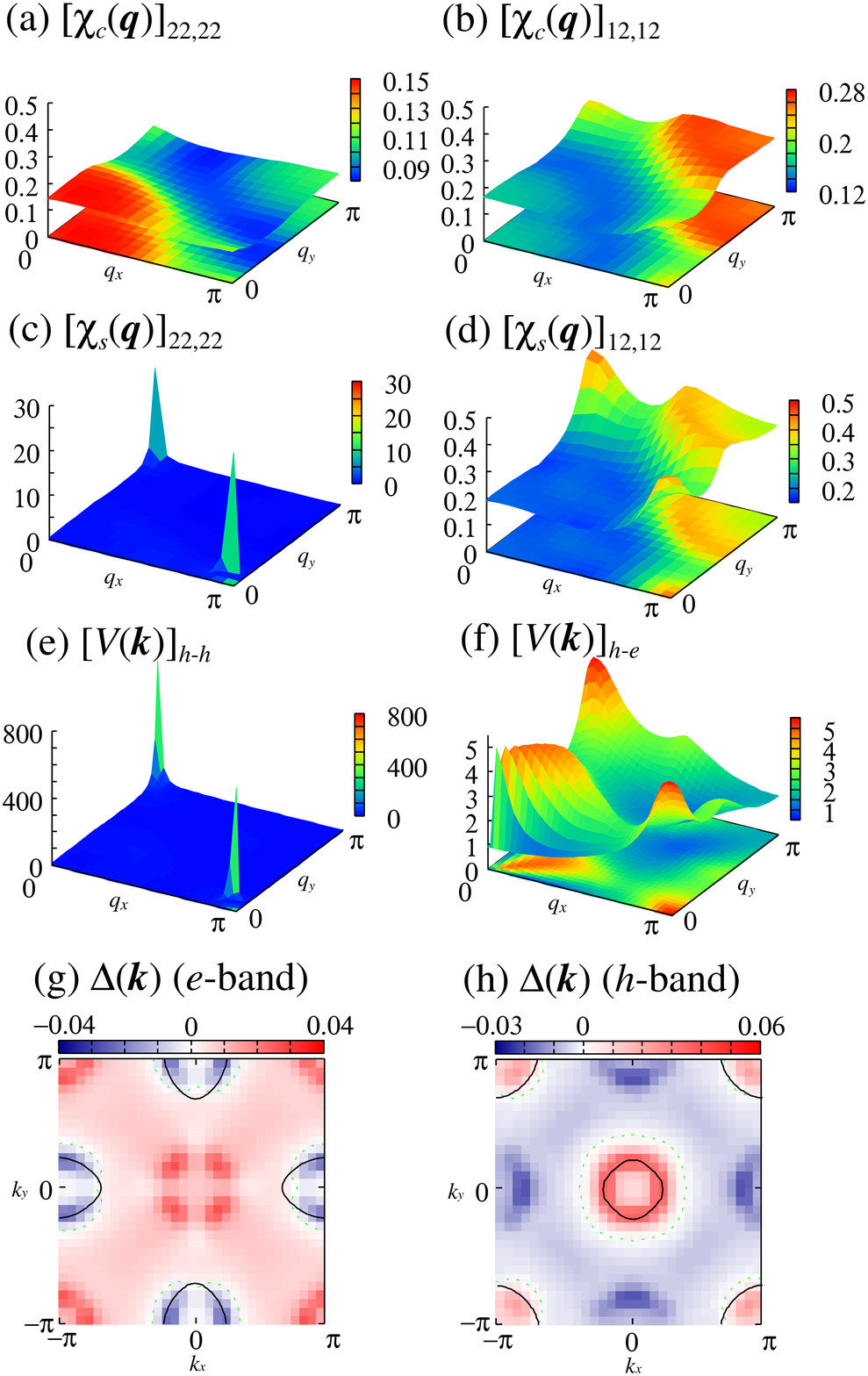}
\caption{(Color online) 
Several components of the charge-orbital susceptibility (a) and (b), those of the spin susceptibility (c) and (d) in the orbital representation, 
the pairing interaction in the hole band (e) and that between the electron and hole bands (f) 
and the gap function in the electron and the hole bands (g) and (h) in the band representation 
as functions of the wave vector with $i\omega_m=0$ for $U'=4$, $J=J'=0.52$, $2g^{2}/\omega_{0}=0$ and $n=2$. 
}\label{fig:Fig6}
\end{center}
\end{figure}

\begin{figure}[t]
\begin{center}
\includegraphics[width=8.0cm]{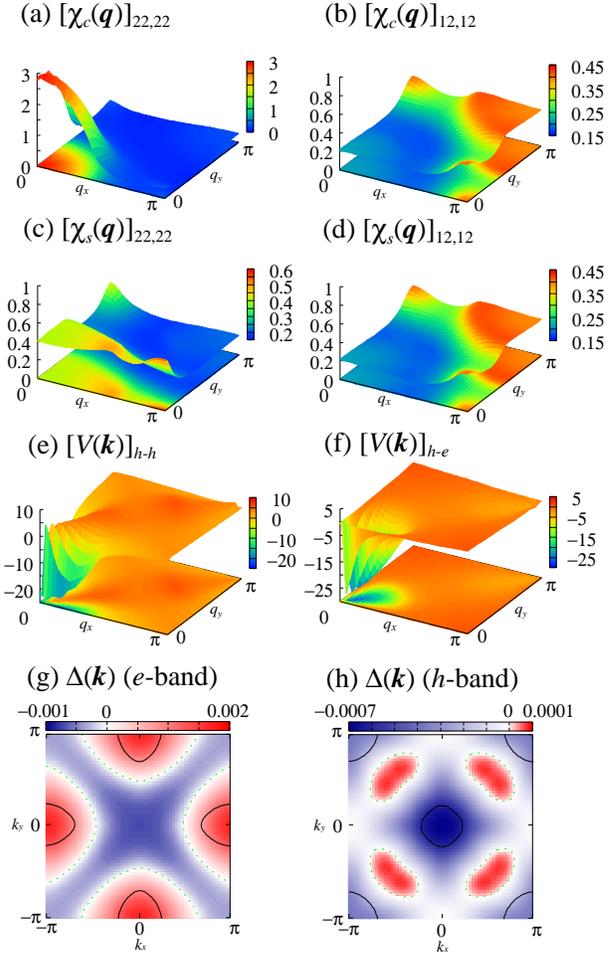}
\caption{(Color online)
Several components of the charge-orbital susceptibility (a) and (b), those of the spin susceptibility (c) and (d) in the orbital representation, 
the pairing interaction in the hole band (e) and that between the electron and hole bands (f) 
and the gap function in the electron and the hole bands (g) and (h) in the band representation 
as functions of the wave vector with $i\omega_m=0$ for $U'=4$, $J=J'=0$, $2g^{2}/\omega_{0}=0.43$ and $n=2$. 
}\label{fig:Fig7}
\end{center}
\end{figure}

As shown in Fig. \ref{fig:Fig5}, the superconductivity is realized near the AFM order, 
where we plot a typical result of the gap function together with the spin and charge-orbital susceptibility and the pairing interaction in Fig. \ref{fig:Fig6}. 
The large repulsive pairing interaction for $\bm{q}\sim(\pi,0)$ due to the AFM fluctuation results in the $s_{\pm}$-wave pairing 
with sign change of the gap function between the hole and the electron FSs\cite{PhysRevLett.101.057003,PhysRevLett.101.087004}. 
It is noted that $\left[\chi_{s(c)}(q_{x},q_{y})\right]_{22,22}$ is not symmetric with respect to the exchange of $q_{x}$ and $q_{y}$ as shown in Figs. \ref{fig:Fig6} (a) and (c) (and also in Figs. \ref{fig:Fig7} (a) and (c)), but $\left[\chi_{s(c)}(q_{x},q_{y})\right]_{11,11}=\left[\chi_{s(c)}(q_{y},q_{x})\right]_{22,22}$ due to the tetragonal symmetry (not shown).

We also observe the superconductivity near the FO order, 
where a typical result of the gap function together with the spin and charge-orbital susceptibility and the pairing interaction is shown in Fig. \ref{fig:Fig7}. 
As shown in Fig.\ref{fig:Fig7} (a), the charge-orbital susceptibility $\left[\chi_{c}(\bm{q})\right]_{22,22}$  (i. e., the longitudinal orbital susceptibility mentioned before) around $\bm{q}=(0,0)$ is largely enhanced by the electron-phonon coupling $g$ between the JT phonon and the longitudinal orbital fluctuation (see eq. (\ref{eq:Hph})). The largely enhanced FO fluctuation induces the large attractive pairing interaction with $\bm{q}\sim(0,0)$ (see Fig.\ref{fig:Fig7} (e)) which dominantly contributes to the superconductivity with the nodeless gap function for both the electron and hole FSs, while the slightly enhanced AFM fluctuation induces the relatively small repulsive pairing interaction with $\bm{q}\sim(\pi,0)$ (see Fig.\ref{fig:Fig7} (f)) which contributes to determine the relative sign of the gap function between the two FSs resulting in the $s_{\pm}$-wave pairing as shown in Figs.\ref{fig:Fig7} (g) and (h). 
This is a striking contrast to the previous result for the 16-band $d$-$p$ model 
where the electron-phonon couplings with the $E_g$ and $B_{1g}$ modes slightly enhance the AFO fluctuation 
and suppress the repulsive pairing interaction due to the AFM fluctuation with $\bm{q}\sim(\pi,0)$, 
and then the $s_{++}$-wave pairing is realized near the FO order due to the FO fluctuation largely enhanced by the JT electron-phonon coupling\cite{JPSJ.79.123707,SSC.152.701}. 
As for the case with the present two-orbital model, 
we also find that the $s_{++}$-wave pairing is realized in the presence of the $B_{1g}$-type JT phonon coupled to the transverse orbital fluctuation, 
in addition to the orthorhombic-type JT phonon coupled to the longitudinal one considered in this study\cite{ishizuka}, 
where the $B_{1g}$-type electron-phonon coupling enhances the AFO fluctuation as similar to the case with 
the previous study mentioned above\cite{JPSJ.79.123707,SSC.152.701}.

Remarkably, the DMFT result of the superconducting region due to the FO fluctuation is expanded as compared to the RPA result, 
while that due to the AFM fluctuation is reduced. 
This is cased by the effect of the local orbital fluctuation which is enhanced 
due to the local correlation effect sufficiently included in the DMFT and results in $V_{\rm loc}<0$ near the FO order as shown in Fig. \ref{fig:Fig5}. 

\subsection{Doping dependence}
\begin{figure}[t]
\begin{center}
\includegraphics[width=6.5cm]{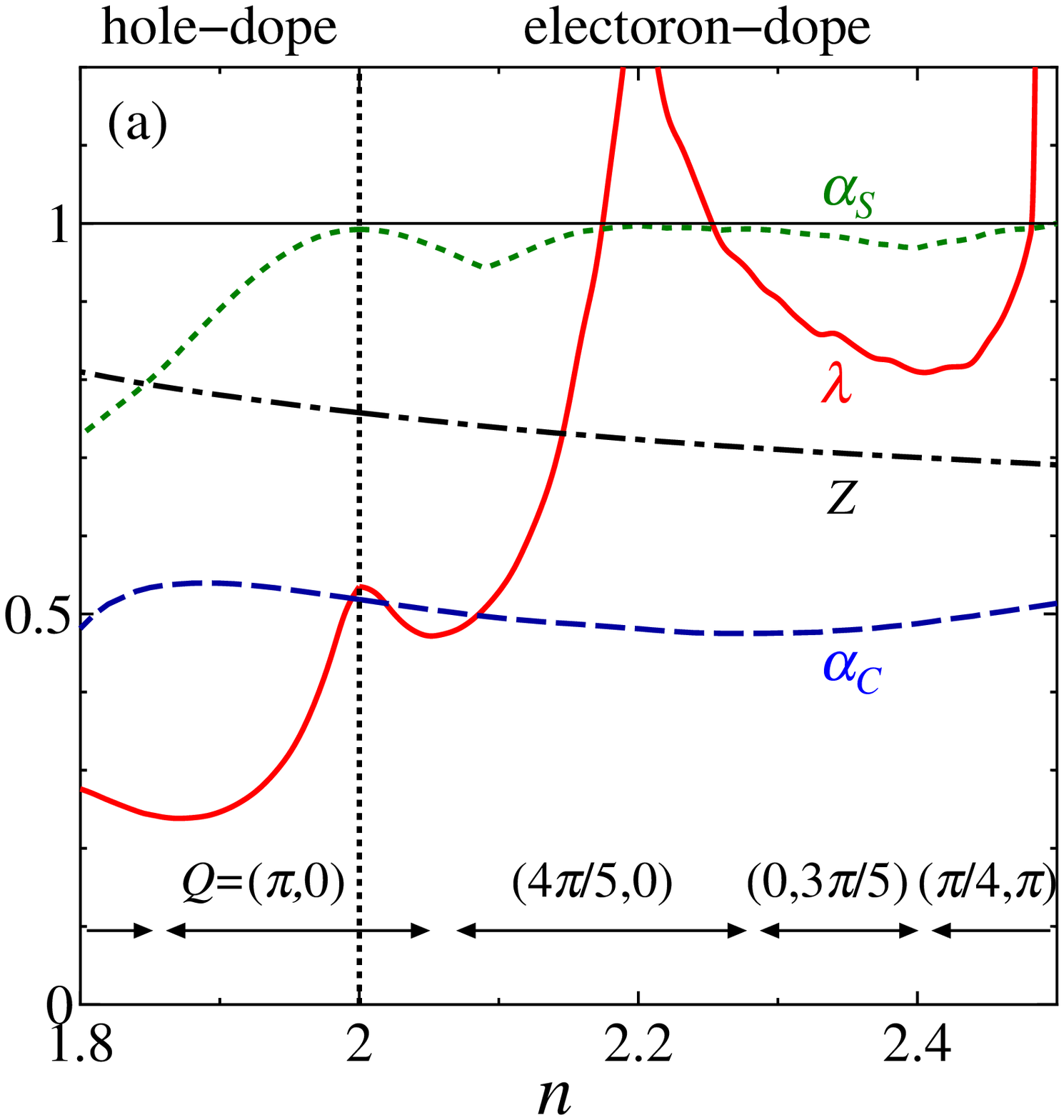}
\includegraphics[width=6.5cm]{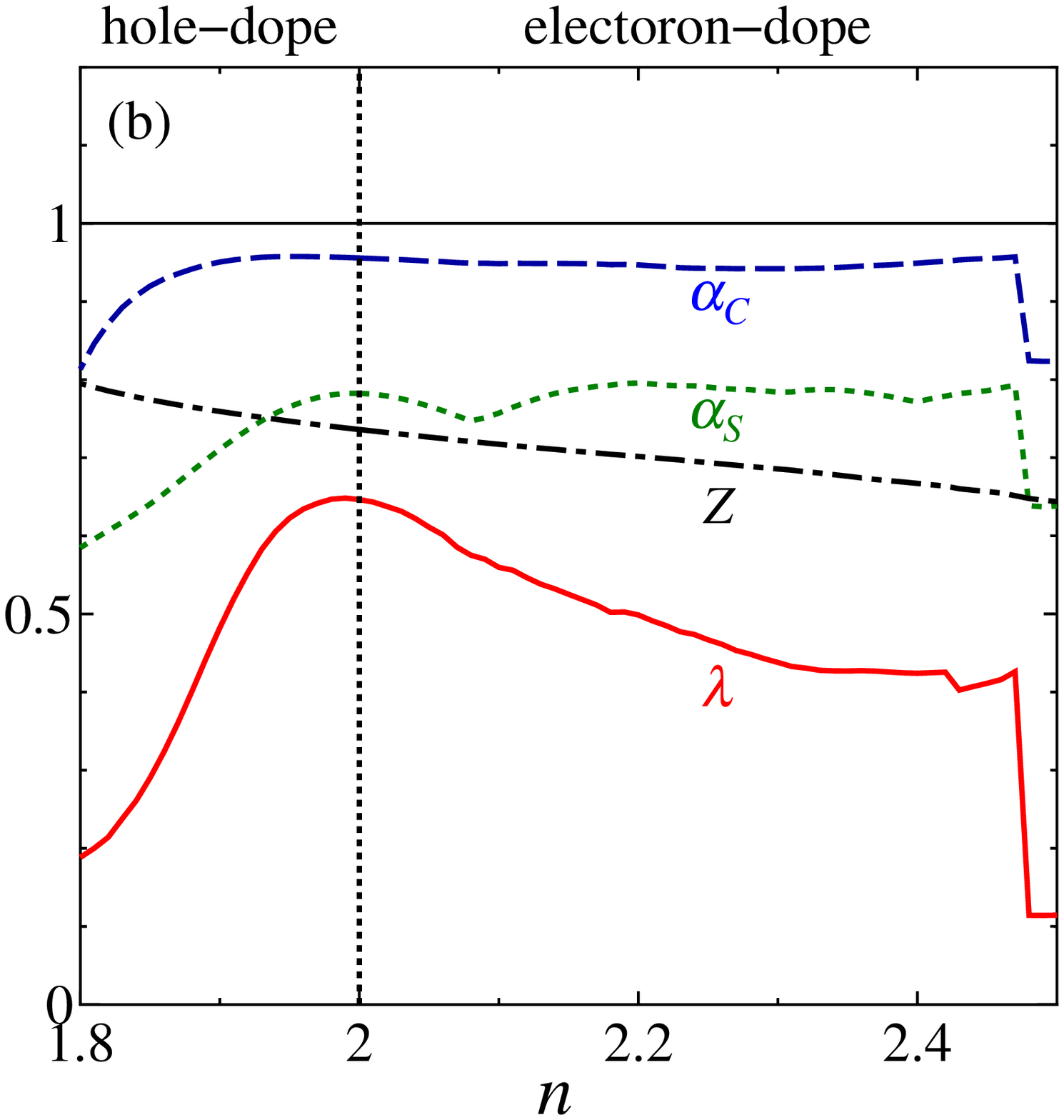}
\caption{(Color online) 
The largest eigenvalues $\alpha_s$, $\alpha_c$ and $\lambda$ 
which reach unity towards the magnetic, charge-orbital and superconducting instabilities 
together with the renormalization factor $Z$ as functions of the electron filling $n$ 
for $U'$=$4$, $J$=$J'$=$0.4$, $2g^{2}/\omega_{0}$=$0.54$ (a) and for $U'$=$4$, $J$=$J'$=$0.2$, $2g^{2}/\omega_{0}$=$0.6$ (b). 
}\label{fig:Fig8}
\end{center}
\end{figure}

Fig. \ref{fig:Fig8} shows the $n$ dependence of the largest eigenvalues $\alpha_s$, $\alpha_c$ and $\lambda$ 
which reach unity towards the magnetic, charge-orbital and superconducting instabilities 
together with the renormalization factor $Z$ in the magnetic fluctuation dominated regime with $\alpha_s>\alpha_c$ (Fig. \ref{fig:Fig8} (a)) 
and in the orbital fluctuation dominated regime with $\alpha_s<\alpha_c$ (Fig. \ref{fig:Fig8} (b)). 
In both regimes, the AFM fluctuation is found to be largely dependent on the electron filling $n$ 
which results in large drastic change in FS nesting responsible for the the strength as well as the wave vector $\bm{q}$ of the enhanced magnetic fluctuation, 
while the FO fluctuation, 
which is enhanced due to the interplay between the electron correlation 
and the JT electron-phonon coupling without the FS nesting effect, is weakly dependent on $n$ 
except for the discontinuous change in the density of states at the Fermi level observed far away from the half-filling. 
Therefore, in the magnetic fluctuation dominated regime, 
the superconductivity due to the AFM fluctuation is largely dependent on $n$ as shown in Fig. \ref{fig:Fig8} (a), 
where $T_c$ is considered to show significant $n$ dependence. 
On the other hand, in the orbital fluctuation dominated regime, 
the superconductivity due to the FO  fluctuation is weakly dependent on $n$ as shown in Fig. \ref{fig:Fig8} (b), 
where $T_c$ is expected to show weak $n$ dependence. 
The latter regime seems to be consistent with the iron-based superconductors 
where the high-$T_c$ superconductivity is observed even in the case with heavily electron-doped compounds as mentioned in $\S$ 1. 

\section{SUMMARY AND DISCUSSIONS}
In summary, we have investigated the electronic state and the superconductivity in the two-orbital Hubbard-Holstein model 
for iron-based superconductors to elucidate the effect of interplay between the electron correlation and the JT electron-phonon coupling 
by using the DMFT+ED method which enables us to sufficiently include the local correlation effects. 
What we have found are as follows: 
(1) In the absence of $J$ and $g$, $\chi_{s}$ and $\chi_{o}$ are equally enhanced due to the effect of $U=U'$. 
(2) In the presence of $J$ and $g$, there are distinct two regimes: for $J \simg 2g^2/\omega_0$, $\chi_{s}$ is enhanced relative to $\chi_{o}$ 
and shows a divergence with $Z\to 0$ at the Mott metal-insulator transition  $J$=$J_c$, 
while for $J \siml 2g^2/\omega_0$, $\chi_{o}$ is enhanced relative to $\chi_{s}$ 
and shows a divergence with $Z\to 0$ at the bipolaronic metal-insulator transition $g$=$g_c$. 
(3) In the former regime, the superconductivity is mediated by the AFM fluctuation enhanced due to the Fermi-surface nesting 
and is largely dependent on the doping, while in the latter regime, the superconductivity is mediated by the FO fluctuation enhanced 
due to the interplay between the electron correlation and the JT electron-phonon coupling without the FS nesting effect and is weakly dependent on the doping. 
The latter regime seems to be consistent with the iron-based superconductors where the high-$T_c$ superconductivity is observed 
even in the case with heavily electron-doped compounds. 

Recently, several authors have investigated the electron correlation effects 
in the two orbital Hubbard model\cite{PhysRevB.79.064517,PhysRevB.84.235115,PhysRevLett.106.186401,EPL.95.17003}. 
The Mott metal-insulator transition has already been discussed within the slave spin mean-field approximation 
and has been found to occur at $U_{c}\sim 2.5W$ with $J$=$0$ and $U_{c}\sim 1.5W$ with $J$=$0.2U$\cite{PhysRevB.84.235115}. 
The results are consistent with the present DMFT results in the absence of the JT electron-phonon coupling. 
The most remarkable point of the present study is the effect of interplay between the electron correlation 
and the JT electron-phonon coupling which is responsible for the FO order and its fluctuation. 
It has been found that, due to the correlation effects beyond the RPA, the FO order is stabilized relative to the AFM order 
and the superconducting region due to the FO fluctuation is expanded while that due to the AFM fluctuation is reduced. 


Experimentally, the iron pnictide superconductors had been observed in the intermediate correlation regime where the band structures from the angle-resolved photoemission spectroscopy (ARPES) are well reproduced by the first-principles band structures by reducing the band width by a factor of $2\sim 3$\cite{AdvPhys.59.803}. However, recent high-resolution ARPES measurements for Ba$_{0.6}$K$_{0.4}$Fe$_2$As$_2$ have  revealed significant orbital dependence of the mass enhancement from 1.3 to 9\cite{JPCM.23.135701}. More recently, remarkable strong correlation effects such as the orbital-selective Mott transition (OSMT) in K$_x$Fe$_{2-y}$Se$_2$\cite{PhysRevLett.110.067003}, where the renormalization factor $Z$ for a specific orbital becomes zero while $Z$ for the other orbitals are finite, and the heavy fermion behavior with the mass enhancement up to 100 in the proximity of the OSMT in KFe$_{2}$As$_2$\cite{PhysRevLett.111.027002} have been observed. To discuss the superconductivity in these systems, we need to fully take into account the strong correlation effects including the Mott transition. In this study, the strong correlation effects including the Mott and bipolaronic transitions have discussed, but, due to simplicity of the present two orbital model, the OSMT has not been included and the superconductivity has been observed far from the Mott (bipolaronic) transition. Quite recently, we have also applied the same approach as the present study to the more realistic five orbital Hubbard model and obtained some preliminary results of the superconductivity near the OSMT\cite{JPSJ.82.123712} which is consistent with the recent experiments mentioned above. However, the effect of the JT electron-phonon coupling, which is considered to play important roles also in the OSMT observed near the FO order and the superconductivity\cite{JPSJ.82.123712}, was not taken into account. Therefore, explicit calculations for such realistic models as a straightforward  but a rather CPU-time consuming extension of the present study are now under way. 

\begin{acknowledgment}
The authors thank Y. Yanagi for useful comments and discussions. 
This work was partially supported by a Grant-in-Aid for Scientific Research from 
the Ministry of Education, Culture, Sports, Science and Technology, 
and also by a Grant-in-Aid for JSPS Fellows. 
\end{acknowledgment}

\bibliography{fe_2orb_JPSJ}

\begin{thebibliography}{10}
\bibitem{notation}
In this paper, $x$, $y$ axes are set to be directed along the nearest
  Fe-Febonds, in contrast to Refs.
  \cite{JPSJ.81.024604,JPSJ.80.073702,JPSJ.79.123707,SSC.152.701} where those
  are second nearest Fe-Fe bonds and the corresponding quadrupole is called as
  $O_{xy}$.
\bibitem{zdependence}
The same approximation has been used in Refs.
  \cite{PhysRevLett.107.137007,PhysRevB.86.064411} for the spin susceptibility
  which is found to well account for the inelastic neutron scattering
  experiment.
\bibitem{gamma}
Within the DMFT, the local self-energy $\hat{\Sigma}$ in eq. (\ref{eq:G}) fully
  includes the effect of the local vertex $\hat{\Gamma}_{s(c)}$ in eq.
  (\ref{eq:gamma}) obtained from the local susceptibility $\hat{\chi}_{s(c)}$
  in eq. (\ref{eq:chiloc}), but the non-local self-energy including the effect
  of the $\bm{q}$-dependent susceptibility $\hat{\chi}_{s(c)}(q)$ in eq.
  (\ref{eq:chi}) is neglected, and then $\hat{\Sigma}$ shows no singularity
  when $\hat{\chi}_{s(c)}(q)$ diverges towards the magnetic (charge-orbital)
  order.
\bibitem{chi}
When $J=J'=g=0$ and $U=U'$, the interaction Hamiltonian eq. (\ref{eq:Hint}) is
  symmetric with respect to the exchange of spin $\sigma$ and orbital $l$, and
  then the relation of $\chi_{s}$=$\chi_{o}$ which is satisfied in the
  noninteracting case as seen in eq. (\ref{eq:chi}) holds also in the case with
  finite $U=U'$.
\bibitem{uvalue}
When $U'=4$, the AFM order takes place even for $J=0$ within the RPA, and then
  we can not discuss the superconductivity due to the AFM fluctuation although
  the superconductivity due to the FO fluctuation is realized for large
  $2g^{2}/\omega_{0}$.
\end{thebibliography}
\end{document}